\documentclass[conference]{IEEEtran}
\usepackage{cite}
\usepackage{amsmath,amssymb,amsfonts}
\usepackage{algorithmic}
\usepackage{algorithm}
\usepackage{graphicx}
\usepackage{textcomp}
\usepackage{xcolor}
\usepackage{booktabs}
\usepackage{multirow}
\usepackage{listings}
\usepackage{tikz}
\usepackage{pgfplots}
\usepackage{hyperref}
\usepackage{colortbl}
\usepackage{array}

\usetikzlibrary{shapes.geometric, arrows.meta, positioning, calc, fit, backgrounds}
\pgfplotsset{compat=1.18}

\definecolor{phase1}{RGB}{52, 152, 219}
\definecolor{phase2}{RGB}{46, 204, 113}
\definecolor{phase3}{RGB}{155, 89, 182}
\definecolor{phase4}{RGB}{231, 76, 60}
\definecolor{phase5}{RGB}{241, 196, 15}
\definecolor{safe}{RGB}{39, 174, 96}
\definecolor{high}{RGB}{192, 57, 43}
\definecolor{medium}{RGB}{230, 126, 34}
\definecolor{low}{RGB}{52, 73, 94}
\definecolor{codebg}{RGB}{248, 248, 248}
\definecolor{keyword}{RGB}{0, 102, 204}

\lstdefinestyle{solidity}{
    backgroundcolor=\color{codebg},
    basicstyle=\ttfamily\footnotesize,
    breaklines=true,
    commentstyle=\color{gray},
    keywordstyle=\color{keyword}\bfseries,
    stringstyle=\color{phase4},
    frame=single,
    framerule=0.5pt,
    rulecolor=\color{gray!50},
    numbers=left,
    numberstyle=\tiny\color{gray},
    numbersep=5pt,
    tabsize=2
}

\def\BibTeX{{\rm B\kern-.05em{\sc i\kern-.025em b}\kern-.08em
    T\kern-.1667em\lower.7ex\hbox{E}\kern-.125emX}}

\begin{document}

\title{A Risk-Stratified Benchmark Dataset for Bad Randomness (SWC-120) Vulnerabilities in Ethereum Smart Contracts}

\author{
\IEEEauthorblockN{1\textsuperscript{st} Hadis Rezaei}
\IEEEauthorblockA{\textit{Department of Computer Science} \\
\textit{University of Salerno}\\
Fisciano, Italy \\
hrezaei@unisa.it}
\and
\IEEEauthorblockN{2\textsuperscript{nd} Rahim Taheri}
\IEEEauthorblockA{\textit{School of Computing} \\
\textit{University of Portsmouth}\\
Portsmouth, United Kingdom \\
rahim.taheri@port.ac.uk}
\and
\IEEEauthorblockN{3\textsuperscript{rd} Francesco Palmieri}
\IEEEauthorblockA{\textit{Department of Computer Science} \\
\textit{University of Salerno}\\
Fisciano, Italy \\
fpalmieri@unisa.it}
}

\maketitle

\begin{abstract}
Many Ethereum smart contracts rely on block attributes such as \texttt{block.timestamp} or \texttt{blockhash} to generate random numbers for applications like lotteries and games. However, these values are predictable and miner-manipulable, creating the Bad Randomness vulnerability (SWC-120) that has led to real-world exploits. Current detection tools identify only simple patterns and fail to verify whether protective modifiers actually guard vulnerable code. A~major obstacle to improving these tools is the lack of large, accurately labeled datasets.

This paper presents a benchmark dataset of 1,752~Ethereum smart contracts with validated Bad Randomness vulnerabilities. We developed a five-phase methodology comprising keyword filtering, pattern matching with 58~regular expressions, risk classification, function-level validation, and context analysis. The function-level validation revealed that 49\%~of contracts initially classified as protected were actually exploitable because modifiers were applied to different functions than those containing vulnerabilities. We classify contracts into four risk levels based on exploitability: \textsc{High\_Risk}~(no protection), \textsc{Medium\_Risk}~(miner-exploitable only), \textsc{Low\_Risk}~(owner-exploitable only), and \textsc{Safe}~(using Chainlink VRF or commit-reveal).

Our dataset is $51\times$ larger than RNVulDet and the first to provide function-level validation and risk stratification. Evaluation of Slither and Mythril revealed significant detection gaps, as both tools identified none of the vulnerable contracts in our sample, indicating limitations in handling complex randomness patterns. The dataset and validation scripts are publicly available to support future research in smart contract security.
\end{abstract}

\begin{IEEEkeywords}
Smart Contract Security, Bad Randomness, Vulnerability Detection, Benchmark Dataset, Ethereum
\end{IEEEkeywords}

\section{Introduction}
\label{sec:intro}

Blockchain platforms have transformed digital transactions through distributed, immutable ledgers. Smart contracts self-executing programs on these platforms automate agreements without intermediaries and have found applications in decentralized finance (DeFi), supply chain management, voting systems, and NFT markets~\cite{atzei2017survey, rezaei2025sok}. However, the immutability of smart contracts creates critical security challenges. Unlike traditional software that can be updated or patched, vulnerabilities in deployed smart contracts cannot be fixed after release~\cite{chu2023survey}. This fundamental limitation has led to significant financial losses. According to OWASP, over \$1.42~billion was lost in 149~documented incidents during 2024~\cite{owasp2025}.

One important but under studied vulnerability is \emph{Weak Sources of Randomness}, catalogued in the Smart Contract Weakness Classification (SWC) registry as SWC-120~\cite{swc120}. Many smart contracts require random numbers for applications such as lotteries, games, and token distributions. Developers commonly use block attributes \texttt{block.timestamp}, \texttt{blockhash}, \texttt{block.difficulty}, or \texttt{block.number} as sources of randomness. However, these values are visible to miners and can be manipulated within protocol constraints, allowing attackers to predict or alter outcomes in their favor~\cite{owasp2023randomness}. The attack on the SmartBillions contract, which resulted in the theft of over 400~ether, exemplifies the exploitation of this vulnerability~\cite{smartbillions2017}.

Several tools have been developed to automatically detect vulnerabilities in smart contracts, including well-known tools such as Mythril~\cite{mueller2018smashing}, Slither~\cite{feist2019slither}, and SmartCheck~\cite{tikhomirov2018smartcheck}. 
These tools employ various techniques including static analysis, symbolic execution, formal verification, and fuzzing. However, empirical studies have demonstrated that these tools suffer from high false positive and false negative rates~\cite{durieux2020empirical}. A~fundamental challenge is that tool developers cannot validate their detectors without high-quality labeled datasets. When datasets mislabel contracts for instance, marking a~contract as ``protected'' simply because an \texttt{onlyOwner} modifier exists somewhere in the code, without verifying whether that modifier actually guards the function containing the vulnerable randomness pattern tools trained on such data learn incorrect detection patterns. This creates a~vicious cycle where inadequate datasets lead to unreliable tools, which in turn produce more inadequate training data~\cite{diangelo2024consolidation}.

The available datasets for the Bad Randomness vulnerability are severely limited. The SWC-120~Registry provides only two example contracts and has not been updated since 2020, with the maintainers acknowledging it may be incomplete and contain errors~\cite{swc120}. SmartBugs Curated~\cite{ferreira2020smartbugs}, which serves as a~widely used benchmark for evaluating detection tools, contains only eight contracts in the Bad Randomness category out of 143~vulnerable contracts with 208~labeled vulnerabilities. RNVulDet~\cite{qian2023demystifying} provides the most comprehensive prior dataset with 34~known vulnerable contracts and 214~carefully audited safe contracts. Our previous work, TaintSentinel~\cite{rezaei2025taintsentinel}, identified 384~vulnerable contracts from the SmartBugs-Wild collection, but limited the search scope to lottery and gaming-related contracts, thereby missing vulnerable contracts in other domains such as DeFi protocols and token distribution systems.

This paper addresses these limitations through three key improvements over prior work. First, unlike TaintSentinel's domain-specific approach, we search across all contract domains using block attribute keywords (\texttt{block.timestamp}, \texttt{blockhash}, etc.) for initial filtering, yielding $5\times$ more vulnerable contracts. Second, we introduce \emph{function-level validation} that traces whether protective modifiers such as \texttt{onlyOwner} are applied directly to functions containing vulnerable code, rather than merely existing elsewhere in the contract. This validation revealed that 49\%~of contracts initially classified as protected were actually exploitable because the mitigation was applied to a~different function than the one containing the vulnerability. Third, we provide \emph{risk-based stratification} that classifies contracts into four levels (\textsc{Safe}, \textsc{Low\_Risk}, \textsc{Medium\_Risk}, \textsc{High\_Risk}) based on exploitability by different attacker types (external, miner, owner) rather than binary vulnerable/safe labels.

\noindent The main contributions of this research are:

\begin{enumerate}
    \item \textbf{Largest SWC-120 benchmark dataset:} We analyzed 17,466~contracts and identified 1,752~vulnerable contracts—$51\times$ larger than RNVulDet (34~vulnerable contracts), $5\times$ larger than TaintSentinel (384~contracts), and $219\times$ larger than SmartBugs Curated (8~contracts).
    
    \item \textbf{Function-level mitigation validation:} We verify that protective modifiers guard the specific functions containing vulnerabilities, not merely exist elsewhere in the contract. This methodology can serve as a~template for improving labeling accuracy in other vulnerability categories.
    
    \item \textbf{Four-level risk classification:} We stratify contracts as \textsc{Safe}, \textsc{Low\_Risk}, \textsc{Medium\_Risk}, or \textsc{High\_Risk} based on which attacker types can exploit them, rather than using binary vulnerable/safe labels that ignore partial mitigations.
    
    \item \textbf{Context-aware analysis:} We distinguish legitimate uses of block attributes (e.g., proof-of-work mining tokens, time tracking) from vulnerable randomness generation (e.g., lotteries, gambling), reducing false positives through semantic understanding.
    
    \item \textbf{Empirical tool evaluation:} We demonstrate that existing detection tools (Slither and Mythril) fail to detect vulnerable contracts with complex patterns, revealing significant gaps in current approaches that focus only on simple modulo operations while missing cases involving \texttt{keccak256} hashing, type casting, and indirect variable usage.
\end{enumerate}

\noindent The remainder of this paper is organized as follows. Section~\ref{sec:background} provides background on bad randomness vulnerabilities. Section~\ref{sec:related} reviews existing detection tools and benchmark datasets. Section~\ref{sec:method} describes our five-phase dataset construction methodology and presents the final dataset. Section~\ref{sec:discussion} evaluates existing detection tools on our dataset and discusses implications and limitations. Section~\ref{sec:conclusion} concludes the paper.

\section{Background}
\label{sec:background}

This section explains bad randomness vulnerabilities in Solidity and common mitigation approaches.

\subsection{Bad Randomness in Solidity}
\label{subsec:bad-random}

Ethereum's execution environment is deterministic by design all nodes must reach the same state given the same inputs. This determinism makes true randomness impossible to generate on-chain. Developers often mistakenly use block attributes as pseudo random sources: \texttt{block.timestamp}, which is the timestamp set by the miner and can be manipulated within approximately 15~seconds according to protocol rules~\cite{soldocs2024}; \texttt{blockhash(block.number)}, which returns the hash of a~specified block where the current block's hash is always zero and only the last 256~blocks are accessible; and \texttt{block.difficulty}/\texttt{block.prevrandao}, which after the Merge returns the RANDAO value that is also manipulable by validators.

A~common vulnerable pattern involves hashing these values to produce a~``random'' number:

\begin{figure}[htbp]
\centering
\begin{tikzpicture}
\node[draw=gray!60, fill=codebg, rounded corners=3pt, inner sep=10pt, text width=0.85\columnwidth] {
\texttt{\textcolor{keyword}{function} random() \textcolor{keyword}{public view returns} (uint) \{}\\[2pt]
\texttt{~~~~\textcolor{keyword}{return} uint(keccak256(abi.encodePacked(}\\[2pt]
\texttt{~~~~~~~~\colorbox{red!20}{block.timestamp}, \colorbox{red!20}{block.difficulty}, msg.sender}\\[2pt]
\texttt{~~~~))) \% 100;}\\[2pt]
\texttt{\}}
};
\end{tikzpicture}
\caption{Vulnerable randomness pattern using predictable block attributes (highlighted).}
\label{fig:vulnerable-code}
\end{figure}

Miners can manipulate these values or selectively publish blocks to influence outcomes. Front-running attacks allow external observers to predict results before transaction execution.

\subsection{Mitigation Mechanisms}
\label{subsec:mitigation}

Several approaches can mitigate bad randomness vulnerabilities. \textbf{Commit-Reveal Schemes} use a~two-phase protocol where participants first submit a~hash of their secret value, then reveal the actual value after all commitments are collected; the combined reveals produce randomness that no single party could predict. \textbf{Chainlink VRF} is a~decentralized oracle providing verifiable random numbers with cryptographic proofs~\cite{chainlinkvrf2024}. For each request, Chainlink VRF generates random values and cryptographic proof of how those values were determined, which is published and verified on-chain. \textbf{Access Control} restricts function access via modifiers such as \texttt{onlyOwner}, limiting who can exploit the weakness; however, this does not eliminate the vulnerability it only reduces the attack surface to trusted parties.

\section{Related Work}
\label{sec:related}

This section reviews existing detection tools and benchmark datasets, highlighting their limitations.

\subsection{Detection Tools}
\label{subsec:tools}

Several static analysis tools detect bad randomness patterns. \textbf{Slither}~\cite{feist2019slither} reports ``weak-prng'' warnings for contracts using block variables in arithmetic operations. \textbf{Mythril}~\cite{mueller2018smashing} uses symbolic execution to detect predictable randomness. \textbf{SmartCheck}~\cite{tikhomirov2018smartcheck} is a~pattern-based analyzer that flags dangerous randomness sources.

\paragraph{Limitations.} These tools operate at contract level and cannot verify whether mitigations actually protect vulnerable functions. They detect only simple patterns (e.g., direct modulo operations) while missing complex cases involving \texttt{keccak256} hashing and type casting.

\subsection{Benchmark Datasets}
\label{subsec:datasets}

Existing benchmark datasets for SWC-120 are summarized in Table~\ref{tab:existing}. The \textbf{SWC~Registry}~\cite{swc120} contains only two example contracts and has not been updated since 2020. \textbf{SmartBugs Curated}~\cite{ferreira2020smartbugs} includes eight contracts in the Bad Randomness category. \textbf{RNVulDet}~\cite{qian2023demystifying} provides 34~vulnerable and 214~safe contracts. \textbf{TaintSentinel}~\cite{rezaei2025taintsentinel} identified 384~vulnerable contracts but limited search to lottery and gaming domains.

\begin{table}[htbp]
\centering
\caption{Existing SWC-120 benchmark datasets and their limitations.}
\label{tab:existing}
\small
\begin{tabular}{@{}lrcc@{}}
\toprule
\textbf{Dataset} & \textbf{Samples} & \textbf{Risk} & \textbf{Valid.} \\
\midrule
SWC Registry~\cite{swc120} & 2 & No & No \\
SmartBugs Curated~\cite{ferreira2020smartbugs} & 8 & No & Partial \\
RNVulDet~\cite{qian2023demystifying} & 34 & No & Unknown \\
TaintSentinel~\cite{rezaei2025taintsentinel} & 384 & No & Yes \\
\textbf{Our Dataset} & \textbf{1,752} & \textbf{Yes} & \textbf{Yes} \\
\bottomrule
\end{tabular}
\end{table}

\paragraph{Limitations.} These datasets are too small for training machine learning models (2-384 samples). None provide risk-level classification or function-level validation. Datasets may label contracts as ``protected'' based solely on modifier presence without verifying the modifier guards the vulnerable function, leading to mislabeled training data.

\section{Dataset Construction and Validation}
\label{sec:method}

This section describes our multi-phase approach to constructing a~labeled benchmark dataset for SWC-120 vulnerabilities. A~key limitation of existing detection tools is their inability to distinguish between \textit{pattern presence} and \textit{actual exploitability}. For example, a~tool may flag a~contract as vulnerable if it uses \texttt{block.timestamp} in randomness generation, then flag it as protected if an \texttt{onlyOwner} modifier exists without verifying whether that modifier guards the specific function containing the vulnerability. This contract-level analysis produces high false positive rates and provides unreliable training data for machine learning models.

\begin{figure}[htbp]
\centering
\includegraphics[width=1.1\columnwidth]{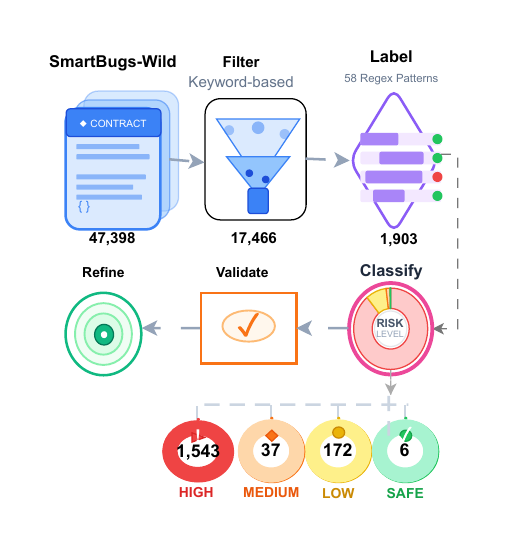}
\caption{Dataset construction pipeline showing contract counts at each phase. The final output is categorized into four risk levels.}
\label{fig:pipeline}
\end{figure}

Our methodology addresses this limitation through five phases: data collection with broad keyword filtering (Section~\ref{subsec:phase1}), vulnerability pattern labeling (Section~\ref{subsec:phase2}), risk-level classification (Section~\ref{subsec:phase3}), function-level validation (Section~\ref{subsec:phase4}), and context-aware refinement (Section~\ref{subsec:phase5}). The function-level validation phase is our core methodological contribution, enabling us to identify contracts where mitigations exist but fail to protect vulnerable code. Figure~\ref{fig:pipeline} illustrates the overall pipeline.

\subsection{Phase 1: Data Collection and Keyword Filtering}
\label{subsec:phase1}

\paragraph{Design Decision.} We use broad keyword filtering rather than domain-specific search (e.g., lottery, gambling) to avoid missing vulnerable contracts in unexpected contexts such as DeFi liquidation mechanisms or token distribution systems. Prior work TaintSentinel~\cite{rezaei2025taintsentinel} limited search to gaming-related contracts, thereby excluding vulnerabilities in other domains.

We used the SmartBugs-Wild dataset~\cite{ferreira2020smartbugs} as our data source, containing 47,398~unique Ethereum smart contracts collected from Etherscan. We filtered contracts containing at least one of the following block attributes: \texttt{block.timestamp}, \texttt{blockhash}, \texttt{block.difficulty}, \texttt{block.number}, \texttt{block.coinbase}, and \texttt{block.gaslimit}. This step reduced the dataset to \textbf{17,466~contracts} (63.1\% reduction). Note that the presence of these keywords does not imply vulnerability; many contracts use block attributes for legitimate purposes such as time tracking. The subsequent phases distinguish between safe and vulnerable usage patterns.

\subsection{Phase 2: Vulnerability Pattern Labeling}
\label{subsec:phase2}

\paragraph{Design Decision.} Existing tools detect only simple patterns (e.g., \texttt{block.timestamp~\%~n}). We developed 58~regular expressions organized into 9~semantic groups to capture complex patterns involving \texttt{keccak256} hashing, type casting, and indirect variable usage that account for the majority of real-world vulnerabilities.

These patterns capture bad randomness sources documented in SWC-120~\cite{swc120}, along with additional patterns identified during analysis such as \texttt{block.prevrandao} and \texttt{gasleft()}. Table~\ref{tab:patterns} summarizes the pattern groups.

Group~G1 detects direct modulo operations with block attributes, such as \texttt{block.timestamp\,\%\,players.length}. Groups~G2 and~G3 address a~common misconception that hashing predictable values provides security: G2~identifies type casting from \texttt{keccak256} or \texttt{sha3} to \texttt{uint}, while G3~detects hashed values used with modulo operations. Groups~G4 and~G5 target \texttt{blockhash} usage, including the deprecated \texttt{block.blockhash} syntax and assignments to variables such as \texttt{answer} or \texttt{result}. Group~G6 matches assignments to variables named \texttt{seed} or \texttt{random} from predictable sources. Group~G7 identifies winner selection patterns using block attributes. Group~G8 detects stored block numbers and direct \texttt{uint} casts from \texttt{blockhash}. Group~G9 captures contextual patterns where randomness-related keywords appear alongside \texttt{keccak256} with block attributes.

\begin{table}[htbp]
\centering
\caption{Vulnerability detection patterns organized by semantic category.}
\label{tab:patterns}
\renewcommand{\arraystretch}{1.1}
\begin{tabular}{@{}clc@{}}
\toprule
\textbf{ID} & \textbf{Pattern Category} & \textbf{Count} \\
\midrule
\rowcolor{phase1!8}
G1 & Direct modulo with block attributes & 10 \\
G2 & Type cast from \texttt{keccak256/sha3} to \texttt{uint} & 11 \\
\rowcolor{phase1!8}
G3 & \texttt{keccak256} hash with modulo operator & 15 \\
G4 & \texttt{keccak256} with \texttt{block.blockhash} & 1 \\
\rowcolor{phase1!8}
G5 & \texttt{blockhash} as answer/comparison & 4 \\
G6 & Seed/random variable with predictable source & 10 \\
\rowcolor{phase1!8}
G7 & Winner selection using block attributes & 2 \\
G8 & Stored block number and uint cast from blockhash & 3 \\
\rowcolor{phase1!8}
G9 & Randomness context with \texttt{keccak256} & 2 \\
\midrule
& \textbf{Total} & \textbf{58} \\
\bottomrule
\end{tabular}
\end{table}

The labeler first checks for Chainlink VRF usage; contracts with VRF are labeled \textsc{Safe}. Remaining contracts are labeled as vulnerable if they match at least one pattern. Of the 17,466~filtered contracts, \textbf{1,903} (10.8\%) matched one or more patterns and proceeded to risk classification.

\subsubsection{Pattern Distribution Analysis}
\label{subsubsec:pattern-dist}

\begin{figure}[htbp]
\centering
\includegraphics[width=1.1\columnwidth]{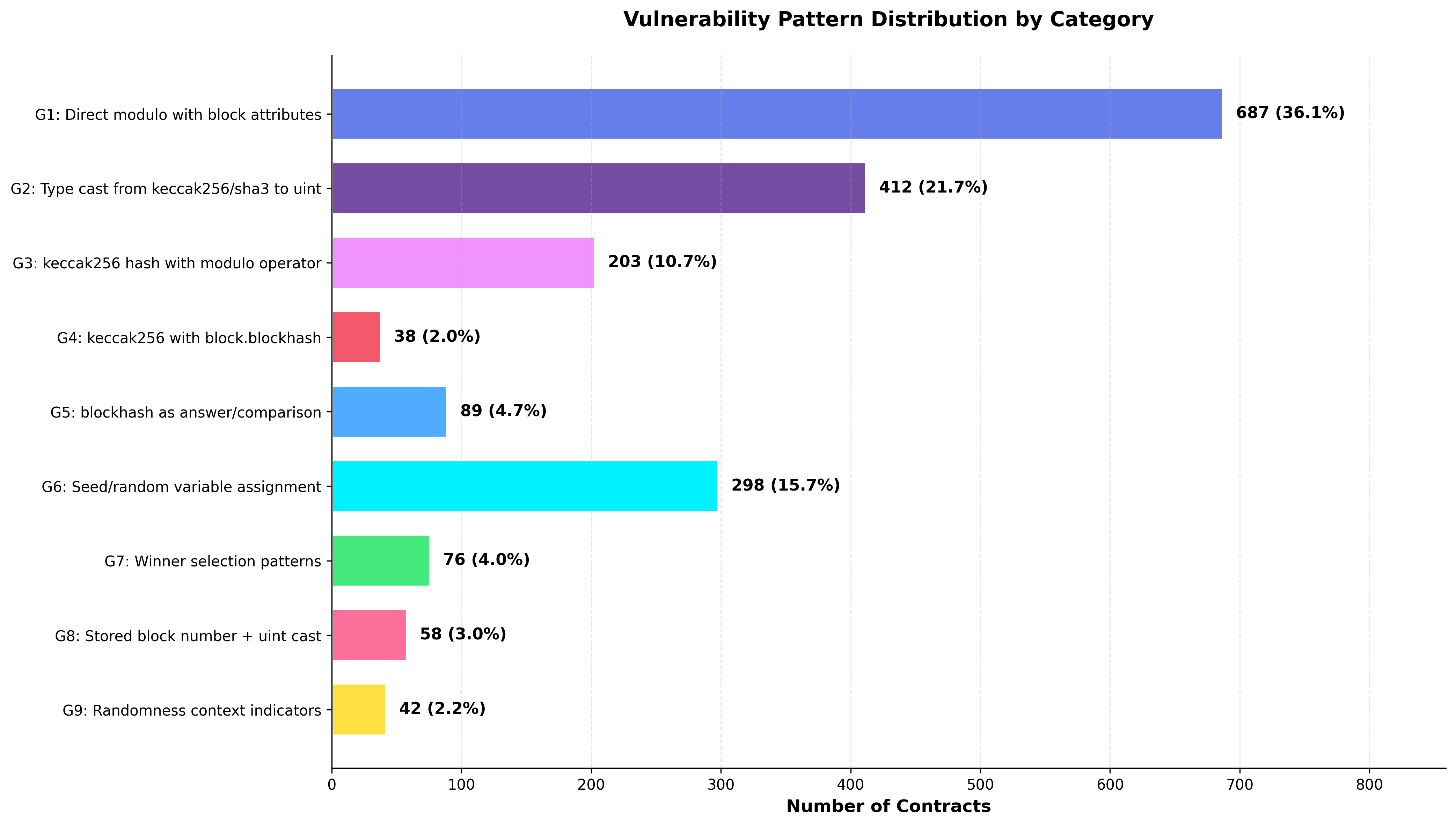}
\caption{Distribution of vulnerability patterns across 1,903~candidate contracts. G1~(direct modulo) and G2~(uint cast) together account for 57.8\% of all patterns.}
\label{fig:pattern-dist}
\end{figure}

Figure~\ref{fig:pattern-dist} shows the distribution of detected patterns across the 1,903~candidate contracts. Group~G1 (direct modulo operations) was the most prevalent, appearing in 687~contracts (36.1\%). Groups~G2 and~G6 followed with 412~(21.7\%) and 298~(15.7\%) contracts, respectively. The high frequency of G1~patterns indicates that developers commonly use the simplest and most vulnerable form of randomness generation. Groups~G4 and~G9 were the least common, each appearing in fewer than 50~contracts.

\subsubsection{Validation with Ground Truth}
\label{subsubsec:ground-truth}

To evaluate the accuracy of our pattern-based labeler, we tested it against a~ground truth dataset compiled from SWC~Registry~\cite{swc120}, SmartBugs Curated~\cite{ferreira2020smartbugs}, and our previous work, TaintSentinel~\cite{rezaei2025taintsentinel}. This dataset includes contracts with documented bad randomness vulnerabilities as well as safe contracts that use block attributes for legitimate purposes such as time tracking.

Table~\ref{tab:ground-truth} presents the evaluation results. The labeler correctly classified all 32~contracts, achieving 100\% accuracy with no false positives or false negatives.

\begin{table}[htbp]
\centering
\caption{Ground truth evaluation results.}
\label{tab:ground-truth}
\renewcommand{\arraystretch}{1.1}
\begin{tabular}{@{}cccc|cccc@{}}
\toprule
\textbf{TP} & \textbf{TN} & \textbf{FP} & \textbf{FN} & \textbf{Acc.} & \textbf{Prec.} & \textbf{Rec.} & \textbf{F1} \\
\midrule
22 & 10 & 0 & 0 & 100\% & 100\% & 100\% & 100\% \\
\bottomrule
\end{tabular}
\end{table}

\subsection{Phase 3: Risk-Level Classification}
\label{subsec:phase3}

\paragraph{Design Decision.} Binary vulnerable/safe labels ignore partial mitigations that restrict attacker capabilities. We classify contracts into four risk levels based on which attacker types can exploit them: external attackers (deploy malicious contracts), miners (manipulate block attributes), and owners (control privileged functions). This stratification enables prioritization of remediation efforts.

Exploitability depends on who can trigger the randomness function and what protective mechanisms are in place. We classify contracts into four risk levels based on detected mitigations and their effectiveness against three attacker types: \emph{external attackers} who deploy malicious contracts, \emph{miners} who manipulate block attributes, and \emph{owners} who control privileged functions. Table~\ref{tab:risk-levels} summarizes the classification criteria and initial results.

\begin{table}[htbp]
\centering
\caption{Risk levels, attacker capabilities, and initial classification results.}
\label{tab:risk-levels}
\renewcommand{\arraystretch}{1.15}
\setlength{\tabcolsep}{4pt}
\begin{tabular}{@{}llcccrc@{}}
\toprule
\textbf{Risk Level} & \textbf{Mitigation} & \textbf{Ext.} & \textbf{Miner} & \textbf{Owner} & \textbf{Count} & \textbf{\%} \\
\midrule
\rowcolor{safe!12}
\textsc{Safe} & VRF / Commit-Reveal & \textcolor{red}{$\times$} & \textcolor{red}{$\times$} & \textcolor{red}{$\times$} & 6 & 0.3 \\
\textsc{Low} & \texttt{onlyOwner} & \textcolor{red}{$\times$} & \textcolor{red}{$\times$} & \textcolor{green!60!black}{$\checkmark$} & 1,167 & 61.3 \\
\rowcolor{medium!12}
\textsc{Medium} & \texttt{tx.origin} / future block & \textcolor{red}{$\times$} & \textcolor{green!60!black}{$\checkmark$} & \textcolor{green!60!black}{$\checkmark$} & 47 & 2.5 \\
\textsc{High} & None & \textcolor{green!60!black}{$\checkmark$} & \textcolor{green!60!black}{$\checkmark$} & \textcolor{green!60!black}{$\checkmark$} & 683 & 35.9 \\
\midrule
\multicolumn{5}{@{}l}{\textbf{Total}} & \textbf{1,903} & \textbf{100} \\
\bottomrule
\end{tabular}
\par\smallskip
\footnotesize{\textcolor{green!60!black}{$\checkmark$}~can exploit \quad \textcolor{red}{$\times$}~cannot exploit}
\end{table}

\paragraph{SAFE.} Contracts that use Chainlink VRF (detected via \texttt{VRFConsumerBase}, \texttt{requestRandomWords}, or \texttt{fulfillRandomness}) or implement a~Commit-Reveal scheme (detected via paired \texttt{commit} and \texttt{reveal} functions). VRF provides cryptographically secure randomness through an external oracle; Commit-Reveal separates commitment from revelation, preventing prediction attacks.

\paragraph{LOW\_RISK.} Contracts that contain access control patterns such as \texttt{onlyOwner}, \texttt{onlyAdmin}, or \texttt{require(msg.sender\,==\,owner)}. External attackers and miners cannot invoke the protected function; exploitation requires the owner to act maliciously.

\paragraph{MEDIUM\_RISK.} Contracts that implement \texttt{tx.origin\,==\,msg.sender} checks or use a~future block pattern (e.g., \texttt{block.number\,+\,delay}). The \texttt{tx.origin} check ensures the caller is an externally owned account, blocking contract-based attacks; miners can still manipulate block attributes within protocol limits.

\paragraph{HIGH\_RISK.} Contracts with no detected mitigation. The randomness function is publicly callable and relies on predictable block attributes. Any attacker can deploy a~contract to predict the outcome within the same transaction.

\begin{figure}[htbp]
\centering
\includegraphics[width=1.1\columnwidth]{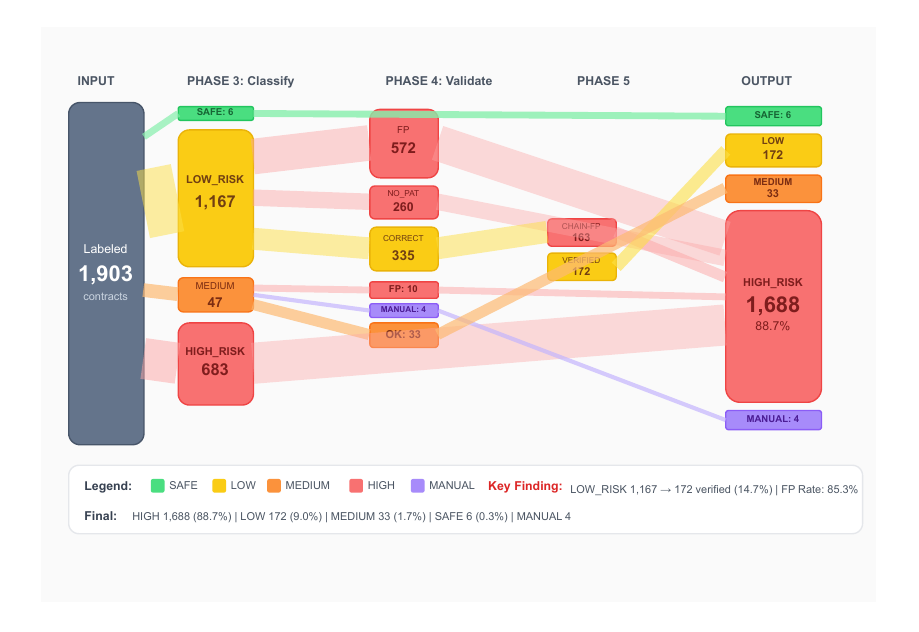}
\caption{Validation flow showing contract reclassification across phases. The \textsc{Low\_Risk} category decreased from 1,167~to 172~contracts (85.3\% false positive rate) after function-level validation.}
\label{fig:sankey}
\end{figure}

\medskip
\noindent The high proportion of \textsc{Low\_Risk} contracts (61.3\%) was unexpected and prompted further investigation. As shown in Figure~\ref{fig:sankey}, subsequent validation phases revealed that many of these classifications were incorrect because the mitigation was not applied to the vulnerable function.

\subsection{Phase 4: Function-Level Validation}
\label{subsec:phase4}

\paragraph{Motivation and Novelty.} Existing tools and datasets check whether a~contract contains both a~vulnerability pattern and a~mitigation mechanism, but do not verify whether the mitigation actually protects the vulnerable code. For example, \texttt{onlyOwner} might protect a~\texttt{withdraw} function while the \texttt{playLottery} function containing bad randomness remains publicly accessible. This contract-level analysis produces false positives where apparently ``protected'' contracts are fully exploitable. Our function-level validation is the first to trace whether mitigations guard specific vulnerable functions, not merely exist elsewhere in the contract.

We developed a~function-level validation algorithm to verify that mitigation mechanisms actually protect the vulnerable code. The algorithm extracts all functions from the contract source code using bracket counting to handle nested structures. For each function, it checks whether any bad randomness pattern from Table~\ref{tab:patterns} is present. If a~vulnerable function is \texttt{public} or \texttt{external}, the algorithm verifies that the mitigation modifier is applied directly to that function. For \texttt{internal} or \texttt{private} functions, the algorithm traces the call chain to identify all public callers and verifies that each caller has the required mitigation.

Algorithm~\ref{alg:validation} provides the formal specification. The algorithm returns one of three verdicts: \texttt{CORRECT} if all vulnerable functions are properly protected, \texttt{FALSE\_POSITIVE} if any vulnerable function lacks protection, or \texttt{NO\_PATTERN\_IN\_FUNCTIONS} if the pattern exists at contract level but not within any function body.

\begin{algorithm}[htbp]
\caption{Function-Level Mitigation Validation}
\label{alg:validation}
\begin{algorithmic}[1]
\REQUIRE Contract source $C$, expected mitigation $M$
\ENSURE Verdict $\in$ \{\textsc{Correct}, \textsc{FalsePositive}, \textsc{NoPattern}\}
\STATE $F \gets \text{ExtractFunctions}(C)$
\STATE $V \gets \{f \in F \mid \text{ContainsBadPattern}(f)\}$
\IF{$V = \emptyset$}
    \RETURN \textsc{NoPattern}
\ENDIF
\FORALL{$f \in V$}
    \IF{$f.\text{visibility} \in \{\texttt{public}, \texttt{external}\}$}
        \STATE $isProtected \gets \text{HasMitigation}(f, M)$
    \ELSE
        \STATE $callers \gets \text{GetPublicCallers}(f, F)$
        \STATE $isProtected \gets \forall c \in callers : \text{HasMitigation}(c, M)$
    \ENDIF
    \IF{$\neg isProtected$}
        \RETURN \textsc{FalsePositive}
    \ENDIF
\ENDFOR
\RETURN \textsc{Correct}
\end{algorithmic}
\end{algorithm}

Applying this algorithm to the 1,167~\textsc{Low\_Risk} contracts revealed significant misclassification. In the first validation pass, 572~contracts (49.0\%) were reclassified as \textsc{False\_Positive} because the \texttt{onlyOwner} modifier was not applied to the function containing the bad randomness pattern. An additional 260~contracts showed \textsc{No\_Pattern\_In\_Functions}, indicating the pattern existed outside function bodies.

For the 335~contracts that passed initial validation, we performed a~second pass with call-chain analysis. This identified 163~additional \textsc{False\_Positive} cases (48.7\%) where an \texttt{internal} function with bad randomness was called by an unprotected \texttt{public} function. After both validation passes, only \textbf{172~contracts} (14.7\% of the original 1,167) were confirmed as correctly classified \textsc{Low\_Risk}.

We applied the same validation to the 47~\textsc{Medium\_Risk} contracts, identifying 10~\textsc{False\_Positive} cases and 4~\textsc{No\_Pattern\_In\_Functions} cases, leaving \textbf{37~confirmed} \textsc{Medium\_Risk} contracts.

\subsection{Phase 5: Context-Aware Refinement}
\label{subsec:phase5}

\paragraph{Design Decision.} Some contracts legitimately use block attributes for purposes other than randomness. Mining tokens use \texttt{blockhash} and \texttt{block.number} for proof-of-work puzzles where predictability does not constitute a~vulnerability ecurity comes from computational cost, not unpredictability. Context analysis prevents false positives by distinguishing intended use.

We performed context analysis on the 260~contracts where patterns appeared outside callable functions (the \textsc{No\_Pattern\_In\_Functions} cases from Section~\ref{subsec:phase4}). The analysis examined keyword frequency across contract names, function names, and variable names to determine the intended use of block attributes.

Contracts containing keywords such as \texttt{mint}, \texttt{difficulty}, \texttt{nonce}, \texttt{mining}, or \texttt{reward} were classified as \textsc{Mining} tokens. Contracts containing keywords such as \texttt{lottery}, \texttt{jackpot}, \texttt{prize}, \texttt{bet}, or \texttt{gamble} were classified as \textsc{Lottery} applications where the bad randomness pattern represents a~genuine vulnerability. Contracts with insufficient context were flagged as \textsc{Unknown} for manual review.

Table~\ref{tab:context-results} shows the classification results. Mining tokens (138~contracts) were excluded from the final dataset because their use of block attributes is not intended for randomness generation. Similarly, 7~contracts from the \textsc{Unknown} category were excluded as they use block attributes solely for time tracking purposes. In total, 145~contracts were excluded as out-of-scope. The remaining contracts were classified as follows: Lottery contracts (108) were labeled as \textsc{High\_Risk}, and 7~\textsc{Unknown} contracts identified as vulnerable gambling applications were also labeled as \textsc{High\_Risk}.

\begin{table}[htbp]
\centering
\caption{Context-aware classification of 260~\textsc{No\_Pattern\_In\_Functions} contracts.}
\label{tab:context-results}
\renewcommand{\arraystretch}{1.1}
\begin{tabular}{@{}lrrl@{}}
\toprule
\textbf{Category} & \textbf{Count} & \textbf{\%} & \textbf{Final Classification} \\
\midrule
\rowcolor{phase5!15}
\textsc{Mining} & 138 & 53.1 & Excluded (not randomness) \\
\textsc{Lottery} & 108 & 41.5 & \textsc{High\_Risk} \\
\rowcolor{phase5!15}
\textsc{Unknown} & 14 & 5.4 & Manual review$^\dagger$ \\
\midrule
\textbf{Total} & \textbf{260} & \textbf{100} & \\
\bottomrule
\end{tabular}
\par\smallskip
\footnotesize{$^\dagger$Result: 7~Excluded (time tracking), 7~\textsc{High\_Risk}}
\end{table}

\subsection{Final Dataset and Comparison}
\label{subsec:final}

After all validation phases, the final labeled dataset contains \textbf{1,758 contracts} distributed as shown in Table~\ref{tab:final-dataset}. The dataset is publicly available on GitHub\footnote{\url{https://github.com/HadisRe/BadRandomness-SWC120-Dataset}}.

 \begin{table}[htbp]
\centering
\caption{Final dataset composition.}
\label{tab:final-dataset}
\renewcommand{\arraystretch}{1.15}
\begin{tabular}{@{}lcrl@{}}
\toprule
\textbf{Category} & \textbf{Count} & \textbf{\%} & \textbf{Label} \\
\midrule
\rowcolor{high!12}
\textsc{High\_Risk} & 1,543 & 87.8 & Vulnerable \\
\textsc{Low\_Risk} & 172 & 9.8 & Vulnerable$^*$ \\
\rowcolor{medium!12}
\textsc{Medium\_Risk} & 37 & 2.1 & Vulnerable$^*$ \\
\textsc{Safe} & 6 & 0.3 & Not Vulnerable \\
\midrule
\textbf{Total} & \textbf{1,758} & \textbf{100} & --- \\
\bottomrule
\end{tabular}
\par\smallskip
\footnotesize{$^*$Vulnerable with partial mitigation (limited attacker scope)}
\end{table}

\subsubsection{Comparison with Existing Datasets}
\label{subsubsec:comparison}

Table~\ref{tab:comparison} compares our dataset with existing SWC-120 benchmark resources. The SWC Registry provides only 2 example contracts for bad randomness. SmartBugs Curated contains 8 contracts in the \texttt{bad\_randomness} category. RNVulDet~\cite{qian2023demystifying}, the most comprehensive prior work, provides 34 contracts with documented vulnerabilities plus 214 contracts labeled as safe.

Our dataset provides \textbf{51$\times$ more vulnerable contracts} than RNVulDet and is the first to include risk-level classification. The function-level validation methodology can serve as a template for improving labeling accuracy in other vulnerability categories.

 \begin{table}[htbp]
\centering
\caption{Comparison with existing SWC-120 benchmark datasets.}
\label{tab:comparison}
\renewcommand{\arraystretch}{1.15}
\begin{tabular}{@{}lrrccc@{}}
\toprule
\textbf{Dataset} & \textbf{Vuln.} & \textbf{Safe} & \textbf{Risk} & \textbf{Func.} & \textbf{Year} \\
\midrule
SWC Registry & 2 & --- & \textcolor{red}{$\times$} & \textcolor{red}{$\times$} & 2018 \\
SmartBugs Curated & 8 & --- & \textcolor{red}{$\times$} & \textcolor{red}{$\times$} & 2020 \\
RNVulDet & 34 & 214 & \textcolor{red}{$\times$} & \textcolor{red}{$\times$} & 2023 \\
\rowcolor{phase2!15}
\textbf{Ours} & \textbf{1,752} & \textbf{6} & \textcolor{green!60!black}{$\checkmark$} & \textcolor{green!60!black}{$\checkmark$} & 2026 \\
\bottomrule
\end{tabular}
\par\smallskip
\footnotesize{Risk = Risk-level classification \quad Func. = Function-level validation}
\end{table}

\section{Discussion}
\label{sec:discussion}
We evaluate existing detection tools on our dataset, discuss implications for tool developers and the research community, analyze the prevalence of unprotected contracts, and acknowledge limitations of our approach.
\subsection{Comparison with Existing Detection Tools}
\label{subsec:tool-comparison}

To evaluate the effectiveness of existing vulnerability detection tools on our dataset, we tested Slither~\cite{feist2019slither} and Mythril~\cite{mueller2018smashing}, two widely-used tools for detecting SWC-120 vulnerabilities (see Section~\ref{subsec:tools}).

\subsubsection{Experimental Setup}

We ran Slither (v0.11.3) on all 1,752 vulnerable contracts in our dataset. For Mythril (v0.24.8), due to its significantly longer analysis time, we evaluated a representative sample of 76 contracts (56 vulnerable, 20 safe). Each tool was configured to detect randomness-related vulnerabilities with a timeout of 120 seconds per contract.

\subsubsection{Results}

Table~\ref{tab:tool-comparison} presents the detection performance of both tools compared to our ground truth labels.

 \begin{table}[htbp]
\centering
\caption{Comparison of existing detection tools on our dataset.}
\label{tab:tool-comparison}
\resizebox{\columnwidth}{!}{%
\begin{tabular}{@{}lrrrrcccc@{}}
\toprule
\textbf{Tool} & \textbf{TP} & \textbf{TN} & \textbf{FP} & \textbf{FN} & \textbf{Acc.} & \textbf{Prec.} & \textbf{Rec.} & \textbf{F1} \\
\midrule
Slither & 0 & 6 & 0 & 1,752 & 0.3\% & 0.0\% & 0.0\% & 0.0\% \\
Mythril & 0 & 5 & 0 & 51 & 8.9\% & 0.0\% & 0.0\% & 0.0\% \\
\midrule
\textbf{Our Labeler} & \textbf{1,752} & \textbf{6} & \textbf{0} & \textbf{0} & \textbf{100\%} & \textbf{100\%} & \textbf{100\%} & \textbf{100\%} \\
\bottomrule
\end{tabular}%
}
\end{table}

Both Slither and Mythril achieved 0\% recall, failing to detect any of the vulnerable contracts in our dataset.

\subsubsection{Analysis of Detection Failure}

To understand why existing tools failed, we analyzed the detection patterns used by each tool:

\begin{itemize}
    \item \textbf{Slither's weak-prng detector} only identifies direct modulo operations on \texttt{block.timestamp}, \texttt{now}, or \texttt{blockhash}. It does not detect patterns involving \texttt{keccak256} hashing, type casting, or indirect usage through variables.
    
    \item \textbf{Mythril's SWC-120 detector} uses symbolic execution to find timestamp dependence but focuses primarily on direct comparisons rather than the complex patterns commonly found in real-world contracts.
\end{itemize}

\begin{figure}[htbp]
\centering
\includegraphics[width=0.8\columnwidth]{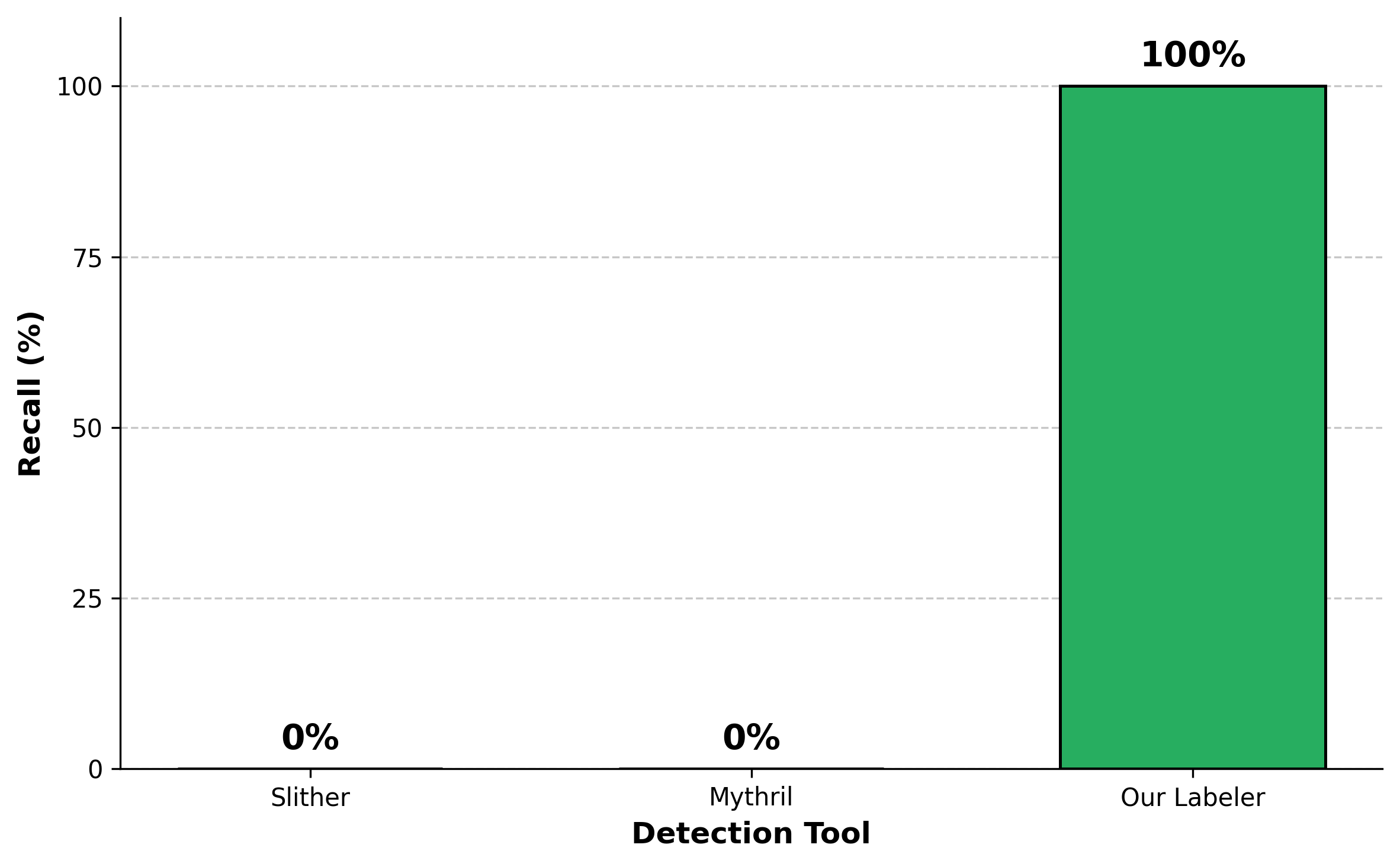}
\caption{Recall comparison of vulnerability detection tools. Existing tools (Slither, Mythril) achieve 0\% recall on our dataset, while our pattern-based labeler achieves 100\% recall by design.}
\label{fig:tool-recall}
\end{figure}

Figure~\ref{fig:tool-recall} illustrates the coverage gap. Our dataset contains 9 pattern groups (G1--G9), but Slither only partially covers G1, and Mythril covers a similarly narrow subset. Groups G2--G9, which account for 63.9\% of our dataset, are completely undetected by both tools.

\subsection{Implications for Detection Tools}
\label{subsec:implications}

Our findings reveal a fundamental limitation in current vulnerability detection tools. Contract-level pattern matching checking whether a contract contains both a vulnerability pattern and a mitigation mechanism produces unacceptably high false positive rates (49\% in our evaluation). Detection tools should implement function-level analysis to verify that mitigations actually protect vulnerable code paths.

\subsection{The Prevalence Problem}
\label{subsec:prevalence}

Our dataset reveals that 87.8\% of contracts with bad randomness patterns have no mitigation whatsoever, classified as \textsc{High\_Risk} in our four-level stratification (contribution~3). This finding was only possible through our function-level validation (contribution~2), which distinguished truly unprotected contracts from those where mitigations exist but fail to guard vulnerable functions.

This prevalence has significant implications for the research community. First, it demonstrates that bad randomness remains a~widespread problem despite years of published research and tool development, indicating a~critical gap between academic knowledge and industry practice. The tools should focus on the 87.8\% of \textsc{High\_Risk} contracts where exploitation requires no special privileges, rather than treating all vulnerabilities equally.

For the developer community, this finding represents substantial financial risk. Many of these unprotected contracts manage funds in DeFi protocols, lotteries, and token distributions where predictable randomness enables direct theft. The high prevalence suggests that developers either remain unaware of this vulnerability class or underestimate its severity.  

\subsection{Limitations}
\label{subsec:limitations}

Our approach has several limitations. First, we analyze only contracts with available source code, excluding bytecode-only contracts. Second, our 58 patterns may not capture all possible bad randomness implementations. Third, we do not trace randomness usage across contract boundaries through inter-contract calls. Fourth, new vulnerability patterns may emerge as Solidity evolves. Finally, Mythril was evaluated on a sample of 76 contracts due to computational constraints; results may vary on the full dataset.

\section{Conclusion}
\label{sec:conclusion}

We presented a benchmark dataset of 1,758 smart contracts labeled for bad randomness (SWC-120) vulnerabilities. Our five-phase methodology data collection, pattern labeling, risk classification, function-level validation, and context-aware refinement addresses critical limitations of existing datasets and detection tools.

The key contributions are: (1) the largest validated SWC-120 benchmark dataset, 51$\times$ larger than RNVulDet; (2) a four-level risk classification based on exploitability; (3) function-level validation revealing that 49\% of apparently protected contracts were actually exploitable; and (4) empirical evidence that existing tools (Slither, Mythril) achieve 0\% recall on diverse bad randomness patterns.

Our dataset is publicly available to support future research in smart contract security analysis and the development of more comprehensive detection tools.

\section*{Acknowledgment}

The authors would like to thank the SmartBugs team for making their dataset publicly available.

\bibliographystyle{IEEEtran}
\bibliography{references}

\end{document}